# Compact magnetic field cycling system with the range from nT to 9.4 T exemplified with $^{13}$C relaxation dispersion and SABRE-SHEATH hyperpolarization

Josh P. Peters[1][*][‡], Charbel D. Assaf[1][‡], Jan-Bernd Hövener[1], Andrey N. Pravdivtsev[1][*]

1. Section Biomedical Imaging (SBMI), Molecular Imaging North Competence Center (MOINCC), Department of Radiology and Neuroradiology, University Hospital Schleswig-Holstein, Kiel University, Am Botanischen Garten 18, 24118 Kiel, Germany

[‡] equal contribution

[*] corresponding authors josh.peters@rad.uni-kiel.de, andrey.pravdivtsev@rad.uni-kiel.de

**Abstract:** We present a compact magnetic field cycling system for high-resolution NMR spectrometers. The system enables the transfer of the sample from $B_0$ field of 9.4 T to ~nT and all fields in between within 1 second. Utilizing a flexible gear rod made the shuttling system more compact, reducing the height to about the height required for filling liquid helium – hence, it can be installed in average-size NMR laboratories (the height of NMR with MFC is only 3.32 m). The system utility is exemplified by measuring $T_1$ relaxation dispersion of the most common liquid state hyperpolarization tracer – [1-$^{13}$C]pyruvate – and magnetic field dependences of signal amplification by reversible exchange enabling alignment transfer to heteronuclei (SABRE-SHEATH) hyperpolarization of [$^{15}$N]pyridine. Using the system, we uncovered the exact relaxation of the pyruvate for a common preclinical dDNP sample composition and gave quantitative estimates for the retained polarization after sample transfer. We modified the observation protocol of SABRE-SHEATH polarization, which, with the high reproducibility of the MFC, provided us with a method to measure the chemical exchange rates of hyperpolarized compounds.

## Introduction

Commercial NMR spectrometers are already versatile and powerful tools for molecular analysis. However, there remains significant room for customization to address specific experimental needs. One such opportunity lies in integrating magnetic field cycling (MFC), which can provide additional information about molecular systems – such as rotational correlation times – that are not always accessible at constant magnetic fields[1–5].

MFC is typically employed to study relaxation behavior as a function of magnetic field strength. This allows one to extract key system parameters, including chemical shift anisotropy (CSA), molecular correlation times, $\tau_c$, and chemical exchange rates, which are otherwise difficult to obtain under static magnetic field conditions[2–4,6–8].

In addition, MFC plays a crucial role in the study of non-equilibrium nuclear spin polarization processes, including chemically induced dynamic nuclear polarization (CIDNP)[9–12], parahydrogen-induced polarization (PHIP)[13–16], Overhauser dynamic nuclear polarization (ODNP)[17,18], and signal amplification by reversible exchange (SABRE)[19–21]. These methods are strongly magnetic-field dependent, and MFC-based studies can reveal properties such as free radical g-factors, signs and magnitudes of hyperfine couplings, and nuclear spin-spin interactions[22]. Moreover, MFC is essential for optimizing polarization transfer conditions in these hyperpolarization techniques[23–25].

Various mechanical designs for MFC systems have been developed, including pneumatic actuators[26], rope-and-pulley systems[27–30], and their combination[31], timing belts[12,32,18], gear rods[33,34], and robotic arms[35,36]. Each has its trade-offs. For instance, rope-and-pulley systems are compact but relatively slow and often rely on gravity unless pneumatically assisted. Gear-rod systems offer simplicity but demand ample vertical clearance due to sealing requirements for the long gear-rod. Systems with powerful floor-mounted motors and timing belts can introduce mechanical instability and vibration, often accompanied by necessitating elevating the NMR magnet and fixing it to the wall, which increases susceptibility to mechanical vibrations. We do not consider MFC systems based on the current electromagnet variation. Such systems are typically optimized for $^1$H detection, and they lack the resolution and sensitivity required for low-γ nuclei like $^{13}$C and $^{15}$N, especially when compared to modern high-resolution NMR systems[1,37–39].

In our work on developing hyperpolarized molecular tracers, we and others have observed unexpectedly rapid relaxation for certain molecules – such as nicotinamide[40,41], succinate[42–44], and fumarate[44] – under specific conditions like low pH or low magnetic fields. In these cases, retained polarization was significantly lower than expected, prompting the development of MFC to study the underlying mechanisms. Moreover, paramagnetic impurities are a known cause of rapid relaxation at low magnetic fields[39], even for long-lived spin states[45,46]. In the same context, MFC was already used to study relaxation as a function of magnetic field (relaxation dispersion) to estimate polarization losses during the transport of hyperpolarized materials[29,38].

To address the needs in relaxometry and hyperpolarization optimization, we developed a compact MFC system to shuttle the sample between high and ultra-low magnetic fields quickly. Although gear-rod systems provide a straightforward solution, their requirement for high vertical clearance makes them incompatible with our laboratory space. At the same time, the compact rope-and-pulley systems were not fast enough to study rapid relaxation. We, therefore, designed an alternative setup utilizing a flexible gear rack to overcome spatial limitations (**Fig. 1**).

This manuscript describes the construction and performance of this flexible gear-rack-based MFC system, which enables sample shuttling between 9.4 T and approximately 10 nT. We demonstrate its utility through $T_1$ relaxation measurements of [1-$^{13}$C]pyruvate under dDNP sample conditions at different magnetic fields (nuclear magnetic relaxation dispersion, NMRD) and estimated losses of polarization during manual transfer of the dDNP hyperpolarized pyruvate to the NMR spectrometer. Moreover, we exemplified its application by hyperpolarizing [$^{15}$N]pyridine using SABRE-SHEATH (signal amplification by reversible exchange in shield enables alignment transfer to heteronuclei)[47,48] and measuring the chemical exchange rate of pyridine with Ir-complex.

## Results and Discussion

Below, we present the design of MFC, automation software, magnetic field profile, and $^{13}$C $T_1$ relaxometry of [1-$^{13}$C]pyruvate and its implication for hyperpolarization experiments and demonstration of $^{15}$N SABRE-SHEATH polarization of [$^{15}$N]pyridine.

### Magnetic field cycling: design and construction

We developed a general-purpose MFC system (**Fig. 1**) using readily available components and custom-designed 3D-printed parts (see the complete list in **Tab. S1**, Supporting Information, SI). The magnetic field setup comprises a top-mounted shelf on the spectrometer and a dedicated MFC frame. The shelf is supported by three brackets that are securely attached to the lifting lugs of the NMR, providing stable alignment and supporting the rails that allow the MFC frame to slide smoothly while ensuring structural stability.

The MFC frame is constructed from four aluminum T-slot profiles (20 mm × 20 mm, 98 cm in length) arranged into a 24 cm × 24 cm square base. These are connected using the same aluminum profiles to form a rigid structure. The first floor of the frame houses the magnetic shield (MS), which is 35 cm tall and rests on a 24 cm × 24 cm plexiglass plate with a thickness of 1 cm. The second floor, with a height of 53 cm, includes the region where the MFC shuttle connects to a flexible gear rack. A second plexiglass plate separates the two floors, providing additional mechanical support and electrical isolation. At the roof of the structure sits the stepper motor and the curved guiding tracks that direct the movement of the gear rack.

The highest point of the NMR spectrometer is 247 cm, with the liquid helium filling port at 242 cm. The helium filling requires 74 cm above the port without any additional space, leading to a needed ceiling height of at least 316 cm or more for the regular NMR operation. Our MFC system sits on NMR at 196 cm level, the top of the MS is 239 cm, and the top of the flexible-gear rod is 332 cm. The total distance from $B_0$ to the beginning of the guiding tube is 120.1 cm. While the MFC sits 16 cm higher than the minimal required ceiling height of the NMR, our MFC is designed to accommodate a shuttle with an NMR tube of up to 56 cm in size (for high-pressure NMR tubes, see below); if only regular NMR tubes are required the shuttles can be as small as 33 cm reducing the total height of the NMR with MFC to 309 cm, around 7 cm below the minimum required ceiling height. This MFC system should fit any room suitable to accommodate an NMR spectrometer.

In contrast, a solid gear rod MFC system would require a total of 120.1 cm of movement plus a shuttle and a stepper motor on top of the MS. This results in at least 392 cm total height to accommodate the longer shuttle, exceeding the minimum required ceiling height by 76 cm and our ceiling height by 45 cm. The design with the shorter shuttle also does not fit in our room. This showcases the need for the flexible rod system presented here.

The length of the shuttle consists of three parts from bottom to top: tube holder, extension, and connector. The tube holder is constructed equal to Bruker's NMR spinner and holds the tube. The extension (middle part) consists of three guiding rods that align the shuttle in the guiding tube. This part must be used as long as the NMR tube is used, possibly including tubings needed for SABRE-based experiments. The top part has an attachment unit to the gear rod. A custom 3D-printed push-click connector is used, enabling rapid shuttle disconnection and sample tube exchange.

As an MS, we chose a compact 4-layer µ-metal shield with integrated shimming coils (MS-1L, Twinleaf). The MS-1L was purchased with custom modifications: the axial bore was increased from 25 mm to 54 mm.

Inside the MS, a hollow anodized aluminum cylinder (outer diameter 30 mm, inner diameter 25.4 mm, and length 420 mm) served as a guide for an NMR tube shuttle and as the core for a resistive solenoid electromagnet (length 300 mm, six layers, 273 turns each, 1 mm isolated copper wire, $R$ = 4.36 Ω).

The other parts were constructed using computer-aided design (CAD) and 3D printing of polylactic acid or poly(methyl methacrylate) (CAD, Inventor, Autodesk, MK3S+, Prusa; Cura slicer, Ultimaker; Formlabs).

A motor drives a flexible gear rack (5217837, RS Components GmbH) to shuttle the sample between the high and low fields. The flexible gear rack slides inside the bent guiding track. At the end of the guiding track, a button is used to automate the calibration of the top position. The construction, while sturdy, is lightweight and easy to assemble. No modification is necessary to the spectrometer, thus making normal operation without interference from the MFC hardware possible. The main construction with the magnetic shield can be removed by loosening four screws and sliding it from the platform if maintenance is needed.

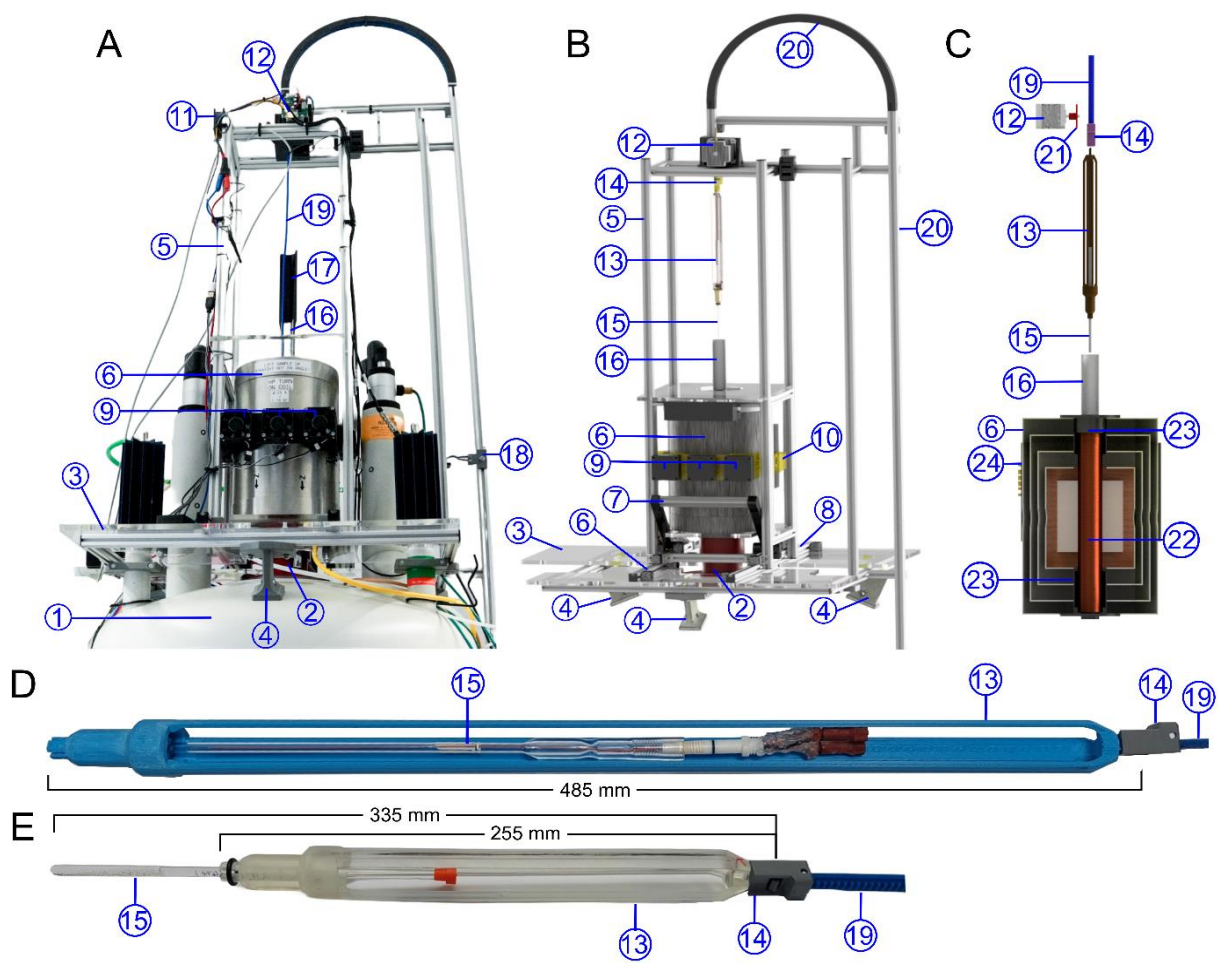

***Fig. 1: Magnetic field cycling (MFC) system.*** *A photo (A) and 3D rendering of the complete (B) and cutout (C) magnetic field cycling (MFC) device atop the spectrometer (NMR, 1) and two used sample shuttles (13) with standard (E) and high-pressure (D) NMR tubes. The MFC enables the cycling of the NMR tubes from $B_0$ magnetic field to the low magnetic field on top of the spectrometer. The MFC frame (5) was mounted on a shelf (3) positioned on top of the 400 MHz spectrometer (1). The shelf is supported by three shelf supporters (4) connected to the lifting lugs of the NMR, ensuring stable alignment. MFC consists of several key components: a magnetic shield (MS, 8), a solenoid electromagnet (SE, 22), and a stepper motor with a spur gear (21). The entire MFC frame (5) with all components slides onto the shelf via integrated rails (6). Once in place, its height could be adjusted using an optional lifter (7) to align it with the NMR guiding tube (2). To maintain thermal stability, the MS (8) and SE (22) are continuously cooled during measurements by circulating air with three cooling fans (9) through dedicated cooling airways (10). An additional cooling fan (11) cools the stepper motor (12) at the top of the structure. An NMR tube (15) is inserted into an MFC shuttle (13), which is secured via a shuttle quick connector (14) to a flexible gear rack (19). This gear rack is inserted into curved guiding tracks (20) and connected to the motor gear (21). The shuttle with the NMR tube travels through the three-component guiding tube system (2, 16, 17). As it enters the magnetic shield (8), the NMR tube passes through SE (22), which is aligned with the MS through dedicated supports (23). Finally, the MS (8) is shimmed via the shimming inputs (24).*

## Magnetic field cycling: electronics and software

To ensure reproducible MFC experiments, it was essential to synchronize all components of the system, which include the 400 MHz NMR spectrometer, the shuttling mechanism, the power supply for the fans used to dissipate heat, the power supplies for the solenoid electromagnet and shimming unit of MS, and when performing SABRE or PHIP experiments a valve system for controlling parahydrogen ($pH_2$) delivery.

The software was designed to incorporate the spectrometer and the MFC software to accurately synchronize motor movements with pulse program (PP) execution. While the main control platform is Bruker's TopSpin software, an auxiliary user program (AU program) executes both the PP and a Python script (communications slave mode, CS) to interface between TopSpin and the MFC software.

Initially, before starting the experiment, the MFC software starts a transmission control protocol (TCP) server to receive commands from the CS, which are used to set parameters such as desired shuttling field, shuttling duration, position, or any other set of variables.

Typically, at least one parameter is varied during the experiment. For example, in relaxometry, the time at a low magnetic field (variable duration parameter, vd) is varied. Such an experiment is convenient for recording data as a pseudo-2D experiment in TopSpin. To accommodate such 2D experiments, up to three arrays and 17 variables can be handed from the AU to the MFC software.

A digital output of the motor is connected to a trigger input on the NMR console to synchronize the motor with the PP. The PP integrates breakpoints to wait for the trigger before proceeding, ensuring accurate timing, especially when running parts of the PP before the shuttling. When working with PHIP, it is possible to use the trigger outputs controlled by the PP to switch the valves for $pH_2$ bubbling.

The MFC software is designed to work modularly: while the base software consists of motor control scripts, standard operation procedures, and many additional modules, the user can program dedicated sequence files. Each PP has a dedicated motor program (MP) file which is dynamically loaded by the MFC software once the corresponding PP is launched. This ensures easy adjustability and maximum flexibility while keeping the MP files lean.

Several additional fundamental scripts were implemented, which convert, for example, a desired magnetic field into position and solenoid electromagnet (SE) current using field-to-position and field-to-coil-current lookup tables. Using this approach, sampling of the field from a few nT to $B_0$ (9.4 T) is easily implemented using field and vd arrays. For example, this approach enabled an entire nuclear magnetic resonance dependency (NMRD) measurement without intervention (see below NMRD curves of pyruvate).

The MFC software uses a serial connection to the motor and constantly monitors the motor's actions, checking for errors such as deviations during the movement. If abnormalities are detected, the sequence is halted to wait for operator input.

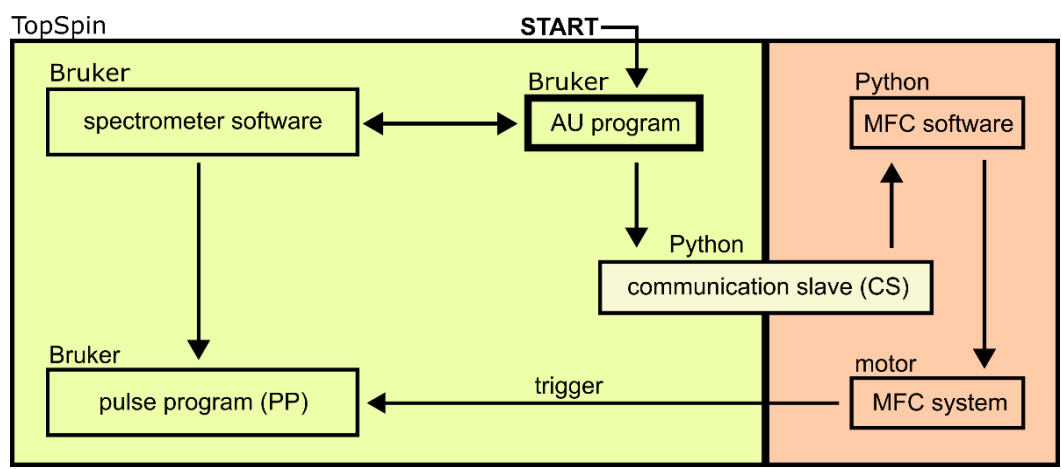

*Fig. 2: Diagram of the communication protocol for controlling the spectrometer and the MFC.* *An AU program plays a central role in the distribution of the tasks and sets up the spectrometer software to run the pulse program (PP) in the desired format. The PP then waits for one or more triggers to conduct the desired excitation and acquisition profile. After setting up the PP, an intermediate Python script (communication slave, CS) is called to communicate with the MFC software and hand over the parameters needed for operation, such as shuttling duration arrays, desired field arrays, and more. The MFC software has a motor program corresponding to every NMR spectrometer PP, which is dynamically loaded and executed. The motor is controlled and constantly monitored. When necessary, an output from the motor can be used to trigger the NMR console and, thus the PP. When an acquisition is completed, it may be repeated with the same variables for averaging or different variables to complete desired pseudo-2D experiments.*

## Magnetic field profile

The magnetic field was measured along the *z*-axis automatically across the sample shuttling distance, starting with the outside position (0%, 0 µsteps) and finishing at the sweet spot of the NMR $B_0$ field (100%, 33,000 µsteps) in 2,000 steps. These measurements were repeated with 31 SE currents between -1.5 and 1.5 A (**Fig. 3**). The magnetic field profile was obtained this way to generate a field-to-position lookup table and corresponding script. Additionally, the magnetic field probe was placed at position 3.34% in the middle of the MS and SE, where the SE current was varied in 1041 steps from 0 to 3 A (about 39.2 W of power dissipation). This calibration was used to calibrate a field-to-current lookup table. When a specific low magnetic field is used, the script is executed and dictates the required current and position for shuttling to access this field using the power supply and the motor.

A zero-field crossing was observable when negative SE currents were used; hence, positive currents were chosen for most future experiments. For fields above 12 mT requiring more than 1.65 A, the MFC system was designed to go to the respective position at the stray field of NMR. For fields below 12 mT, the MFC instead is going to the position 3.34%, and the power supplies of SE and MS will set the required field. Hence, the shuttling duration and polarization losses during sample transfer to fields above 12 mT will be lower, aiding in sensitivity when working with quickly relaxing molecules. However, the homogeneity of the SE field is higher compared to the field in the NMR bore and was thus preferred for < 12 mT.

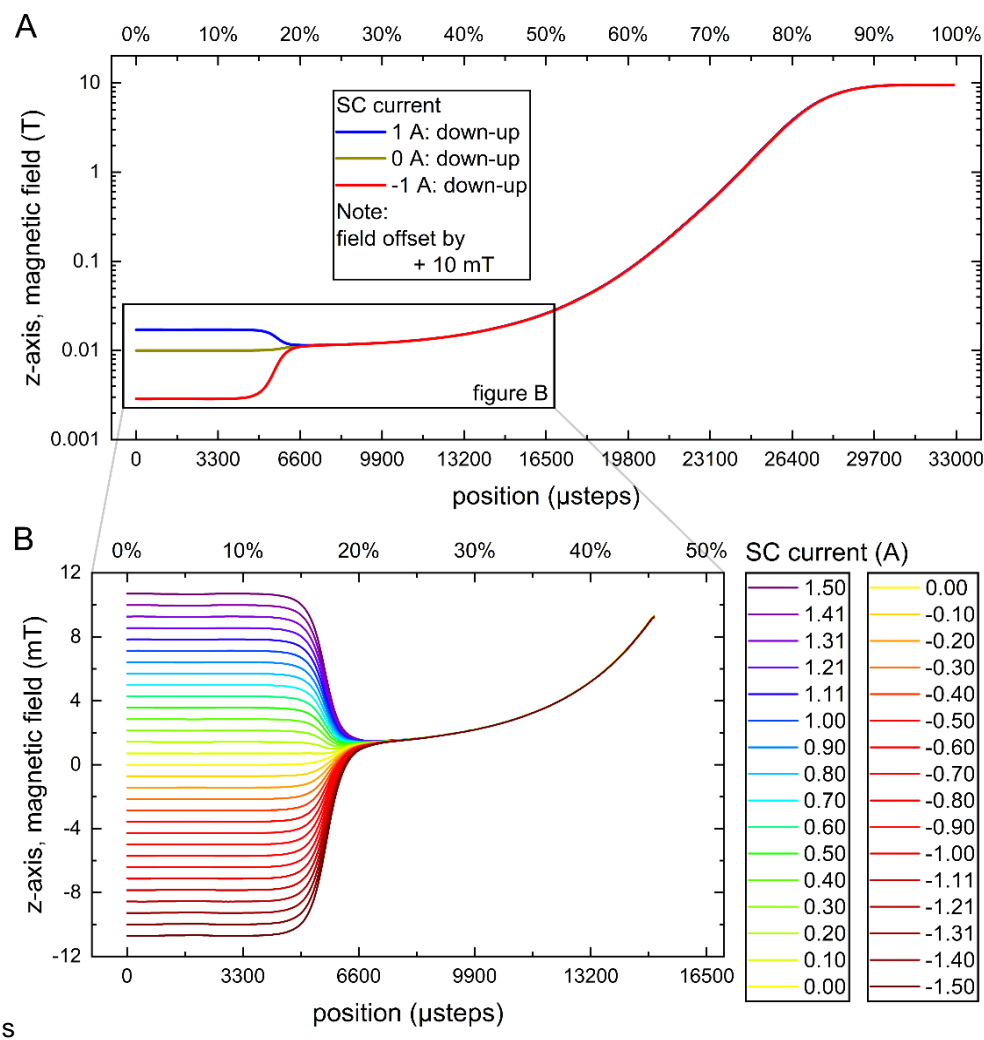

*Fig. 3: Magnetic field profile.* Complete magnetic field profile of the MFC (A) and a section of the profile with the varied current in the solenoid electromagnet (SE) inside the magnetic shield (MS) (B). (A) A plateau of the magnetic field can be observed close to the sweet spot of $B_0$ field at 33,000 μsteps position, and a steep increase when moving between 15,000 and 28,000 μsteps. Note that the field is given with the +10 mT offset to incorporate negative fields in a log-y axis and demonstrate that the field orientation changes when the negative field is applied. (B) The field profile as a function of position and current supplied to SE. The field is homogenous in at least 4,000 μsteps or ~ 64 mm (0.0364 mm/μstep). The field to current ratio of the SE is ~7.1 mT/A. Note that the exact magnetic field at 0 A of applied current depends on the shimming of the MS, but may reach a few nT.

## MFC spectra

INEPT was found to have a substantial gain of 2.53 in signal when used for pyruvate (**Fig. 4A**), while shortening the relaxation time at $B_0$ with $^1$H, $T_1^{1H}$=4.9 s, compared to $^{13}$C, $T_1^{13C}$=45.5 s. This decreased the measurement time to achieve a similar sensitivity using normal acquisition by 59.4 times and was chosen for all subsequent non-SABRE experiments. The INEPT spectrum was obtained during calibration of the INEPT delays with active decoupling, which led to a slight drift in chemical shift due to temperature changes.

To assess the quality of the spectra after MFC, the spectra before and after shuttling were compared (**Fig. 4B**). A slight loss of signal can be observed after shuttling the sample (position 100% - 3.34% - 100% - 100 ms settling delay - acquisition), due to relaxation; however, no distortion of the lineshape was visible. Slight distortions are observable in $^1$H spectra when line broadening is not applied. However, our system is currently optimized for $^{13}$C and $^{15}$N spectroscopy, and a longer settling delay may be chosen if needed for $^1$H observation which is a more sensitive nucleus to distortions.

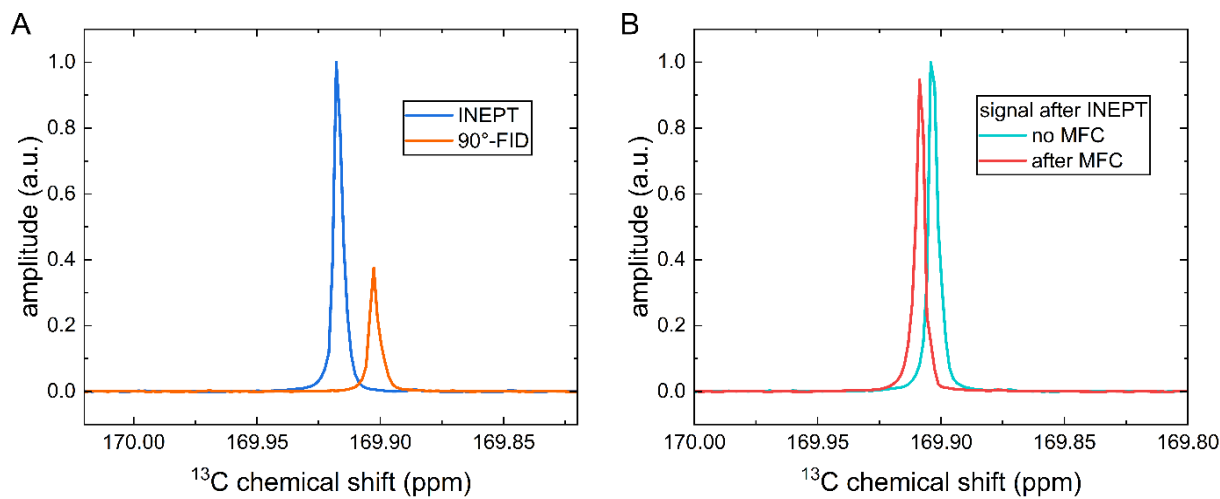

***Fig. 4: Stationary and MFC $^{13}$C NMR spectra of [1-$^{13}$C]pyruvate.*** *(A) Comparison of [1-$^{13}$C]pyruvate spectra with and without $^1$H-$^{13}$C INEPT enhancement at $B_0$. INEPT increased the signal by a factor of 2.53. This enhancement, combined with a shorter relaxation time of $^1$H, $T_1^{1H}$=4.9 s, compared to $^{13}$C, $T_1^{13C}$=45.5 s, led to a 59.4 times reduction of measurement time, providing the same SNR. (B) $^1$H-$^{13}$C INEPT enhanced NMR spectra without (stationary) and after MFC in the low magnetic field. No distortions in the line shape were visible for the spectrum measured 100 ms after MFC. All spectra were acquired with $^1$H decoupling, and the line broadening of 0.2 Hz was applied. Notice the different resonance shift in (A) due to temperature swing when performing multiple INEPT sequences with decoupling for INEPT calibration, which were not present to this extent when longer repetition times were used (B).*

## NMRD of [1-$^{13}$C]pyruvate

To determine the relaxation behavior for the most common dDNP hyperpolarized tracer, [1-$^{13}$C]pyruvate, we examined its NMRD in a composition typical for dDNP after an actual dissolution DNP experiment containing 36 mM Trizma buffer, 45 mM NaCl, 0.24 mM EDTA, 45 mM NaOH, and 151 µM trityl AH111501 radical in 90% H$_2$O and 10% D$_2$O (**Fig. 5**). The sample contained 10% D$_2$O for locking and shimming. No purification was used after dissolution.

For the DNP sample, contaminants such as trityl radical and possible paramagnetic impurities from the overheated dissolution module may be present. However, the low-field $T_1$ of 40 s was still much longer when compared to more quickly relaxing hyperpolarized[40–44], showcasing why pyruvate performs robustly in standard dDNP experiments: it does not require elevated magnetic fields for transport and is relatively resilient to stable radicals. The maximum $T_1$ was reached at 3.1 T with 55.2 s and dropped to 43.2 s at the highest field of 9.4 T.

In the dDNP experiments performed with this sample before, the $T_1$ values were 71.2, 77.3, and 57.8 s for the fields of 0.55, 1.05, and 9.4 T. The liquid state polarization was calculated to be 27.3% after 19.5 s transfer. All hyperpolarization observed $T_1$ values were larger compared to the values measured for a thermally polarized sample (**Fig. 5A**) at room temperature: 52.4, 53.9, and 43.2 s measured at 0.51, 1.02, and 9.4 T, respectively, using MFC at room temperature. Since the sample from the dDNP had elevated temperatures, and temperature was found to increase $T_1$[49], that may account for the observed differences.

## [1-$^{13}$C]pyruvate polarization losses during transfer

Using the relaxation data obtained with the MFC, we calculated the polarization losses during the transfer from the pyruvate polarizer to the measurement site. To determine the relaxation profile, we measured the magnetic field profile along the transfer path. In our case, it was from the dDNP to the NMR system. The exact timing of the sample transfer – including dissolution, distribution into a smaller tube, actual transport, and insertion into the NMR spectrometer – was estimated, and the total duration was set to 19.5 seconds, as in our typical experiments. This yielded the magnetic field profile experienced by the sample during the transfer, shown in **Fig. 5A**. Based on this profile, the polarization loss (from an initial value set to 100%) was calculated for three different scenarios: (i) no additional magnets, (ii) transfer within a dedicated magnet between the dDNP and NMR systems, and (iii) a dissolution receiver vessel equipped with an additional magnet.

A clear difference between the no-magnet and transfer-magnet cases can be observed, 6% more retained of the initial polarization for the dDNP sample, which is 10.1% more than without transfer magnets. The difference between cases (ii) and (iii) was only 0.6%. However, it may provide a bigger relevance for quickly relaxing molecules such as [1-$^{15}$N]nicotinamide[40,41].

For the dDNP sample (**Fig. 5A**), no transfer magnet was used after dDNP, and the observed polarization was 27.3%. Using the calculated polarization losses of 38.1% from the initial value (**Fig. 6**), the initial value can be estimated to be about 44.1% at the time of dissolution.

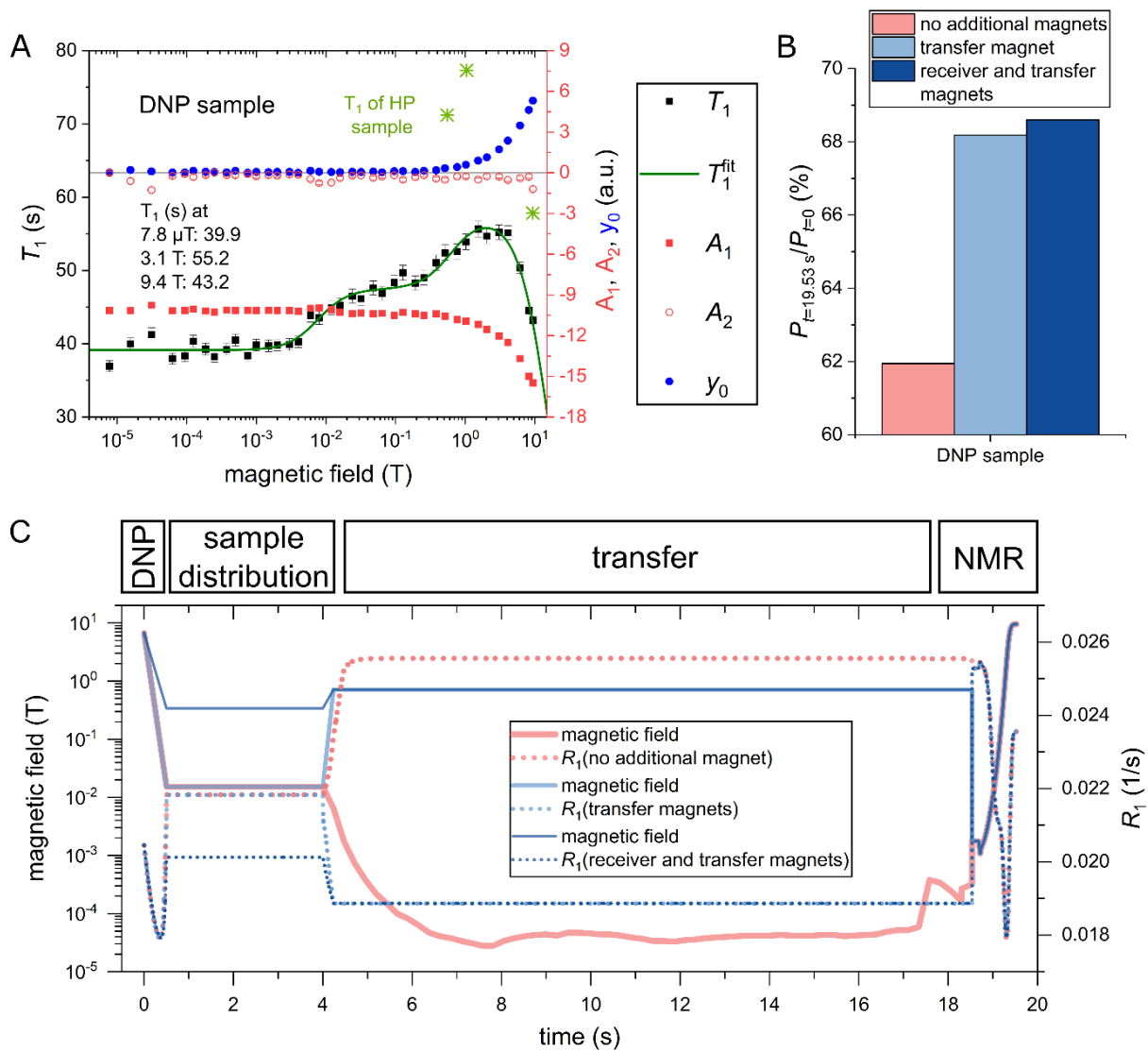

*Fig. 5: NMRD of [1-$^{13}$C]pyruvate at fields from 7.8 µT to 9.4 T (A) and impact of [1-$^{13}$C]pyruvate relaxation and magnetic field on the obtained polarization (B) when transferring from the polarizer to the NMR spectrometer (C).* (A) A significant difference between the low-field $T_1$ and high-field $T_1$ values, showcasing the need for MFC to quantify the changes in relaxation accurately. The field around 1-3 T gave the longest relaxation time across all samples: the $T_1$ was always longer than 50 s, in this range, while at Earth's magnetic fields, it could fall below 30 s.
(B) One can see a clear difference between the three transfer cases: no additional magnets (red), transfer magnet (light blue), and receiver and transfer magnets (darker blue). This originates from the different $T_1$ at low to medium fields. The dDNP sample would retain only 62.0% without additional magnets. The transfer magnet retained about 6% more of the initial polarization. The benefit of an additional receiver magnet were limited. The initial polarization value is set to 100%.
(C) The three field profiles during transfer from the polarizer to the NMR have been measured to estimate the polarization losses as a result of sample transfer: corresponding relaxation rates are given with dashed lines.
<u>Methods</u>: The measured kinetics were fit with a biexponential decay function: $A_1 e^{-t/\tau_1} + A_2 e^{-t/\tau_2} + y_0$. The first value with the largest corresponding amplitude $A_1$ (red squares) was associated with $T_1$ (black squares). $y_0$ (blue circles) is associated with thermal polarization in the given field. The second time constant $\tau_2$ has been shared across all fields for each sample individually to compensate for the observed nonexponential decay; note that the *contribution is relatively low (amplitude $A_2$, red circles) and 0 in the Tris sample. The green line corresponds with the fit values of a relaxation model as described in the methods; the fit parameters can be found in **Tab. 1**. $T_1$ of the hyperpolarized (HP) sample is added using green stars. Sample: 89 mM [1-$^{13}$C]pyruvate with pH of 7.7 of a standard dDNP sample after actual dissolution DNP experiment containing 36 mM Trizma buffer, 45 mM NaCl, 0.24 mM EDTA, 45 mM NaOH, and 151 µM trityl AH111501rasdical in 90% $H_2O$ and 10% $D_2O$. Pyruvate pKa is about 2.5[50].*

## MFC for SABRE-SHEATH

In SABRE, the target substrate is polarized upon transient interaction of $pH_2$ and substrate with Ir-complex (**Fig. 7A**). The optimal polarization transfer field (PTF) varies across substrates due to specific SABRE matching conditions, which are determined by the *J*-couplings and the differences of Larmor frequencies of the hydride and target nuclei[21,47]. Typically, the polarization in such experiments is most effective at a particular level anti-crossing (LAC) field; however, in the case of SABRE, the actual optimal PTF is shifted due to chemical exchange-induced modulations of spin interactions.[51,52] For $^{15}$N SABRE-SHEATH, the optimal polarization transfer occurs in the microtesla (µT) field range. Magnetic field sweeps were performed to identify the optimal PTF for pyridine (**Fig. 7B**).

The PTF dependence of $^{15}$N hyperpolarization was studied by sweeping the field between –3 and +4 µT. The maximum signal for pyridine was observed at a PTF of +0.6 µT (**Fig. 7B**). The positive PTF is parallel to the $B_0$, whereas the negative PTF is antiparallel. In our setup, we observed that negative PTFs led to reduced or even inverted polarization. A similar effect was found before for a similar MFC system[53] and was attributed to the partial polarization transfer during shuttling through the zero field, where polarization transfer continues and negatively affects the attained polarization level.

To prevent this inversion, we repeated the SABRE-SHEATH experiment with an additional magnetic field ramp to +20 µT after $pH_2$ bubbling was stopped and before mechanical sample transfer (**Fig. 7B**). This procedure ensured the instantaneous projection of the spin order on the state at a relatively high field away from possible polarization transfer fields. Hence, further sample transfer did not contribute to the polarization redistribution, as the zero-field regime was avoided. The maximum achieved polarization was 0.4%.

Note the partial H-D exchange on pyruvate with methanol (pyruvate-methanol exchange), which leads to the resonance corresponding to three isotopomers (**Fig. 7C**). Considering this and previous observation of D-H exchange when deuterated pyruvate was used[54,55] in deuterated methanol (pyruvate-$H_2$ exchange), and H-D exchange between $H_2$ and methanol ($H_2$-methanol exchange), one can conclude that protons (or deuterons) of $H_2$, methanol, and ortho protons (carbon positions 2 and 6) of pyridine are all loosely connected and experience mutual exchange.

After optimized SABRE-SHEATH polarization at a low magnetic field, the exchange rates between free and bound pyridine could be measured at high magnetic fields, leveraging high $^{15}$N polarization. Following SABRE-SHEATH and MFC, a frequency-selective 180° pulse was applied to one form of pyruvate (either free or equatorially bound), followed by an evolution delay $\tau_e$ and a 90° hard pulse for detection at high field. This pulse sequence enables tracking of the polarization transfer due to chemical exchange, allowing us to quantify the exchange kinetics in both directions, from free to bound pyridine and from bound to free (**Fig. 7D**). The obtained effective exchange rate constant *k = (5.77 ± 0.19) s$^{-1}$ was obtained* at 288 K. Considering concentration of bound and free pyruvate the dissociation rate constant was estimated to be $k_d$ = *(9.5 ± 0.3) s$^{-1}$* that is within error margin coincides with the previously measured values using INEPT enhanced[56].

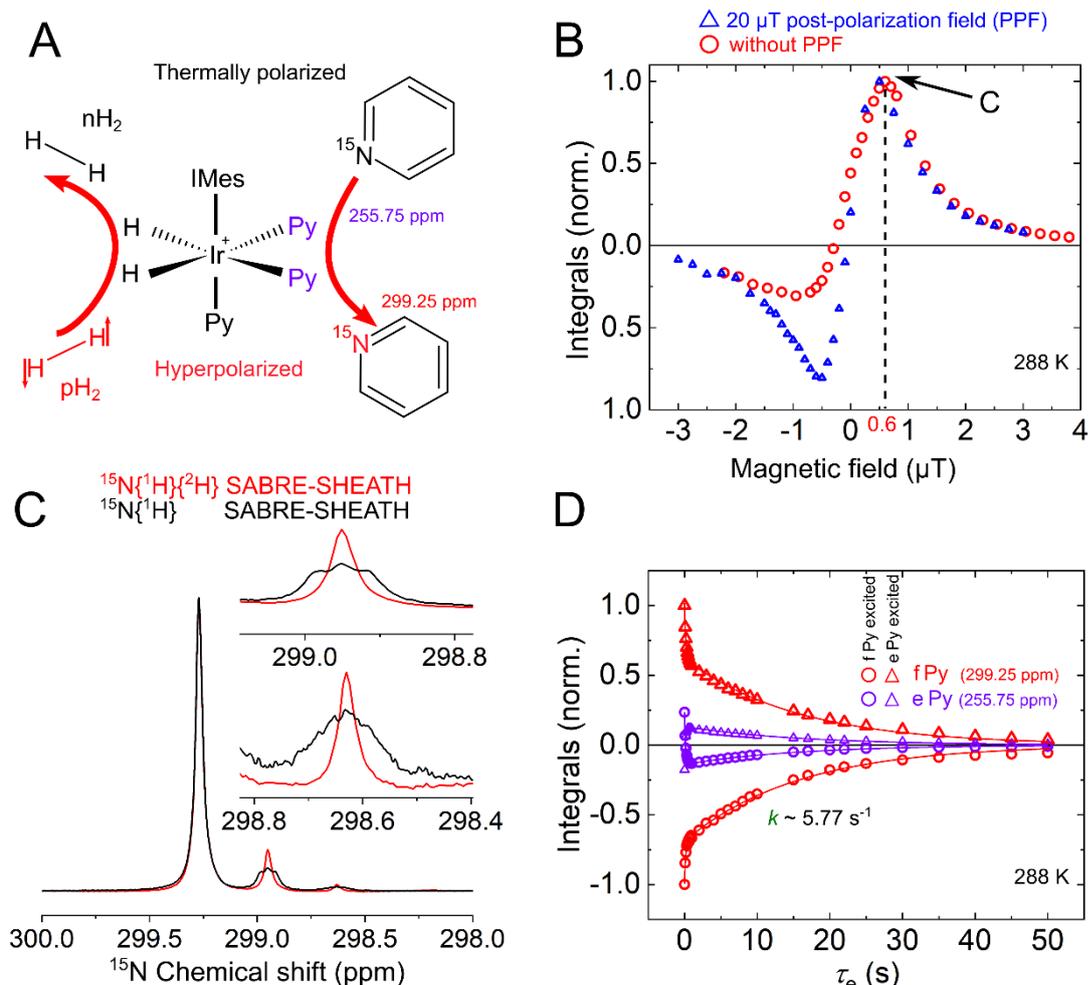

***Fig. 7: Hyperpolarization of [$^{15}$N]pyridine with SABRE-SHEATH.*** *(A) Scheme of SABRE-SHEATH hyperpolarization of pyridine, pH$_2$, and pyridine exchanges with the IrIMes complex that can result in hyperpolarization of $^{15}$N. (B) Normalized integrals of $^{15}$N signals of free pyridine after SABRE-SHEATH as a function of the magnetic field with (blue triangles) and without (red circles) 20 µT post-polarization field (PPF). (C) $^{15}$N{$^1$H}{$^2$H} (red) and $^{15}$N{$^1$H} (black) SABRE-SHEATH spectra showing polarized free pyridine formed by supplying 8.5 bar of pH$_2$ in methanol-d$_4$ at 288 K. Note that the pyridine experienced partial deuteration; therefore, three spectra are present: pyridine-h$_5$, -d,h$_4$, and -d$_2$,h$_4$. (D) Chemical exchange between free and bound pyridine was measured at high magnetic fields and 288 K after SABRE-SHEATH hyperpolarization using selective polarization inversion of free (triangles) or bound (circles) pyridine. The integrals of free pyridine are red, and those of bound pyridine are violet. Kinetics were fitted with a shared biexponential decay function, resulting in two decay rates of k = (5.77 ± 0.19) s$^{-1}$ and R = (0.064 ± 0.0008) s$^{-1}$.*

## Conclusion

We demonstrated only some of the use cases of the MFC system, focusing on the hyperpolarization applications. The availability of the NMRD field dependencies could help one design the most optimal hyperpolarization experiment, such that the critical steps of purification or sample transfer are carried out at the fields with the longest relaxation time. Pyruvate is not very sensitive to the paramagnetic stable radicals when it is in an aqueous buffered solution with EDTA, a typical dDNP sample composition. This explains the high polarization values reported for dDNP even when the radical is not filtered from the solution. The fields between 0.1 T and 3 T provided the highest $^{13}$C $T_1$ of [1-$^{13}$C]pyruvate; hence, it should be used when possible for manipulations with the hyperpolarized pyruvate or during transfer using a magnetic tunnel or transfer magnet. Some other molecules experience rapid polarization losses at low fields and pH[40–44]. The investigations of corresponding NMRD could shed light on their relaxation mechanism, how this rapid relaxation can be circumvented, e.g. using the CIDER effect, and at what conditions the relaxation time reaches maximum.

SABRE hyperpolarization can be performed at high and low magnetic fields. Hyperpolarization at low magnetic fields typically provides superior hyperpolarization yield because it enables one to accumulate polarization on the free substrate[23,57,58]. The high-field experiments, however, due to high resolution, allow the study of the chemical interactions[56,59]. Therefore, combining low the fields, as was demonstrated here and before[48] with exchange rate measurements, can bring about a valuable synergy.

# Methods

### INEPT enhanced NMRD experiment

The INEPT sequence[60] with refocusing was used without changes and without phase cycling:
$^{13}$C: -----$\tau$-180$_X$-90$_X$-$\tau$-180$_X$-$\tau$-90$_Y$-(shuttling)-90$_X$-FID
$^{1}$H: 90$_X$-$\tau$-180$_X$-90$_Y$-$\tau$-180$_X$-$\tau$------(shuttling)-decoupling
The optimum INEPT interpulse delay, $\tau$, was calibrated for each sample experimentally. Subsequent MFC experiment followed the following sequence: (1) $^{1}$H thermal relaxation at $B_0$ for $D_{B0}$ = 19 s, (2) INEPT sequence to transfer $^{1}$H polarization to longitudinal polarization of [1-$^{13}$C] of pyruvate, (3) shuttling to desired low-field, either inside NMR bore or in SE with appropriately set current, (4) waiting for $^{13}$C relaxation at low-field for $D_{LF}$, (5) shuttling to $B_0$ followed by settling delay of 100 ms, (6) $^{13}$C 90° excitation and signal acquisition with $^{1}$H decoupling.

### SABRE-SHEATH experiment

A typical SABRE-SHEATH experiment to study polarization as a function of the polarization field ($B_{pol}$) involved the following steps: (1) before proceeding with and repeating the following steps, the MS was shimmed to reach a homogenous zero field (0-100 nT variation in magnetic field magnitude across 3 cm), (2) thermal relaxation and temperature equilibration at $B_0$ for $D_{B0}$ of 15 s, (3) the NMR tube was shuttled to the isocenter of the MS, where the desired magnetic field was set (ranging from −3 µT to +4 µT); (4) pH$_2$ was bubbled through the solution for $t_b$ period; (5) optionally a +20 µT post-polarization field (PPF) was switched on; (6) shuttling to $B_0$ followed by a field-cycling settling delay of 100 ms; and (7) $^{15}$N 90° excitation and signal acquisition with optional $^{1}$H and/or $^{2}$H decoupling.
This procedure was repeated either by altering $B_{pol}$ and fixing $t_b$, or by fixing $B_{pol}$ and altering $t_b$.

### SABRE sample preparation and pH$_2$ delivery

We conducted SABRE experiments using 50 mM [$^{15}$N]pyridine (486183, Sigma-Aldrich) and 4 mM of the iridium N-heterocyclic carbene complex [Ir(COD)(IMes)Cl] ([Ir], where COD = 1,5-cyclooctadiene and IMes = 1,3-bis(2,4,6-trimethylphenyl)imidazol-2-ylidene)[61] (obtained from University of York), dissolved in methanol-$d_4$ (MeOD, Sigma-Aldrich). 92% parahydrogen (pH$_2$) was prepared and supplied at 8.5 bar to a 5 mm high-pressure NMR tube (522-PV-7, Wilmad-LabGlass) containing 400 µL of the SABRE solution. The NMR tube was connected to the pH$_2$ control unit[62] via two flexible 1/16″ polytetrafluoroethylene (PTFE) tubes (1528XL, IDEX Health & Science LLC). These tubes were threaded through the cap of the NMR tube and secured using fast glue (Loctite 3090). For the experiments, the NMR tube was inserted into a custom NMR tube carrier (**Fig. 1D**), which was then loaded into the MFC system. This setup ensured that the supply and exhaust lines remained connected throughout the MFC process, allowing for continuous or on-demand pH$_2$ bubbling. To avoid mechanical obstruction during shuttling, the PTFE tubes were stuck to the quick connector body (top of the shuttle, **Fig. 1**) with slight tension, minimizing the risk of jamming.

### Exchange rates obtained with SABRE-SHEATH

After preparing SABER-SHEATH polarization, the sample was transferred to the high field of 9.4 T for detection. To measure the chemical exchange between free and bound pyridine under SABRE-SHEATH conditions, a frequency-selective inversion pulse followed by a variable delay was introduced before the final read-out 90° broadband excitation pulse. A frequency-selective inversion 180° pulse (Gaus1_180r.1000 shape, 0.0079983 W, 20.79 dB) was applied for 10 ms to either the free or bound pyridine resonance, followed by a variable free evolution interval $\tau_e$ from 0 to 50 s and a 90° hard pulse (75 W, 21.25 µs). The experiment was performed twice: (1) inverting the polarization of free pyridine, and vice versa, (2) inverting the polarization of bound pyridine and monitoring polarization distribution between two sites. Data were fitted using a global fit of a biexponential decay function: $A_1 \times exp(-\tau_e \times k) + A_2 \times exp(-\tau_e \times R)$ simultaneously for both kinetics. Here $k$ is an effective exchange rate and $R$ is effective relaxation[63,64,56].

Considering initial concentrations of [Ir] = 4 mM and [Pyridine] = 50 mM, leading to the concentration of free pyridine of 50-3·4 mM = 38 mM. The dissociation rate constant of pyruvate, $k_\text{d}$, can be then estimated as follows[64,56]: $k_\text{d} = \frac{k}{0.5 + \frac{[\text{Ir}]}{[\text{free Pyridine}]}} \cong k \cdot 1.65$.

## NMRD fitting

The fitting was conducted using a model that included scalar relaxation of the second kind ($R_J$, Eq. 2[65]) and chemical shift anisotropy (CSA) induced relaxation ($R_\text{CSA}$, Eq. 4 in the ESI of Ref. [66]).

The functions for $R_J$ was

$$R_J = \frac{8\pi^2 J^2}{3} S_\text{X}(S_\text{X} + 1) \frac{\tau_C}{1+(\omega_\text{X} - \omega_{13_\text{C}})^2 \tau_C^2}, \qquad \text{Eq. 1}$$

where $J$ is the fitted scalar coupling constant, $S_\text{X}$ is the spin number of the coupled nucleus, $\tau_C$ is the correlation time characteristic of this interaction, $\omega_\text{X}$ and $\omega_{13_\text{C}}$ are the Larmor angular frequencies of the scalarly coupled nuclei, where X is $^1$H or $^2$H, depending on the solvent.

The function for $R_\text{CSA}$ was

$$R_\text{CSA} = \frac{1}{5}(\omega_{13_\text{C}} \delta_\text{CSA})^2 \frac{\tau_\text{CSA}}{1+(\omega_{13_\text{C}} \tau_\text{CSA})^2}, \qquad \text{Eq. 2}$$

where $\delta_\text{CSA}$ is the CSA of the studied nucleus and $\tau_\text{CSA}$ corresponding correlation time.
The NMRD had two elbows at fields below 1 T, which results in the necessity of having two functions like $R_J$. The final fitting curve was achieved by fitting the following superposition:
$$R = R_{J1} + R_{J2} + R_\text{CSA}. \qquad \text{Eq. 3}$$
Possibly, the two J are the consequence of proton exchange at the COOH group and conversion between ketone and geminal diol (hydrate) form. The fitting resulted in too large values for $J$, unreasonable for typical molecules (**Tab. 1**), indicating different original mechanisms, such as dipole-dipole interaction, which has similar field dependency, but $J$ is substituted with the corresponding dipole-dipole term[2]. Hence, the actual mechanism cannot be deduced solely from NMRD.

## Relaxation of hyperpolarization

To estimate the remaining hyperpolarization, we solved the Bloch equation without thermal magnetization, assuming it is much less than the observed polarization value and SNR:

$$\frac{P(T)}{P(0)} = \exp\left(-\int_0^T \frac{dt}{T_1(B(t))}\right) = \exp(-R_1^\text{avg} T) \qquad \text{eq 5}$$

Where $T$ is the total time of the sample transfer, $T_1(B(t))$ is the relaxation time as a function of magnetic field, which depends on time and $R_1^\text{avg}$ is the average $T_1$ relaxation rate constant experienced by the sample during the transfer.

*Tab 1. Fitted parameters of Fig. 5.*

| Sample | DNP | st. dev. |
|---|---|---|
| $J_1$ (Hz) | 17.8 | 5.7 |
| $J_2$ (Hz) | 147.5 | 18.9 |
| $\tau_{C1}$ (ns) | 715.8 | 177.5 |
| $\tau_{C2}$ (ns) | 8.6 | 5.0 |
| $\delta_\text{CSA}$ (ppm) | 45.8 | 12260 |
| $\tau_\text{CSA}$ (ps) | 3.6 | 1925 |

## Supporting Information

Additional materials, list of of the shelf components to assemble MFC system are available online (.pdf)
All the raw data will be available on Zenodo repository.

## Author Information

Josh Peters, ORCID 0000-0003-1019-4067;

Charbel D. Assaf, ORCID 0000-0003-1968-2112;

Jan-Bernd Hövener: ORCID 0000-0001-7255-7252;

Andrey N. Pravdivtsev, ORCID 0000-0002-8763-617X

## Author Contributions

ANP, JBH: conceptualization; ANP, CA, JP: investigation; ANP, JP, CA: analysis, writing – original draft; ANP, JBH, JP, CA: writing, final versions; CA: design and construction; JP: electronics and software programming; ANP, JBH: supervision, funding acquisition. All authors contributed to discussions and interpretation of the results and have approved the final version of the manuscript.

## Acknowledgements

We acknowledge funding from the German Federal Ministry of Education and Research (BMBF, 03WIR6208A hyperquant), DFG (555951950, 527469039, 469366436, HO-4602/2-2, HO-4602/3, GRK2154-2019, EXC2167, FOR5042, TRR287). MOIN CC was founded by a grant from the European Regional Development Fund (ERDF) and the Zukunftsprogramm Wirtschaft of Schleswig-Holstein (Project no. 122-09-053). We acknowledge the financial support of Kiel University through validation funds and the assistance of the fabrication center "FabLab" for their support in the design and construction of the MFC, in particular, Jan Kirchner.

**Keywords:** magnetic field cyclin • hyperpolarization • relaxation • SABER • chemical exchange • nuclear magnetic resonance

## References

1. Koenig, S. H. & Brown, R. D. Field-cycling relaxometry of protein solutions and tissue: Implications for MRI. *Prog. Nucl. Magn. Reson. Spectrosc.* **22**, 487–567 (1990).
2. Kimmich, R. *NMR: Tomography, Diffusometry, Relaxometry*. (Springer-Verlag Berlin Heidelberg, 1997).

3. Bodurka, J., Seitter, R.-O., Kimmich, R. & Gutsze, A. Field-cycling nuclear magnetic resonance relaxometry of molecular dynamics at biological interfaces in eye lenses: The Lévy walk mechanism. *J. Chem. Phys.* **107**, 5621–5624 (1997).

4. Bertini, I. *et al.* NMR Spectroscopic Detection of Protein Protons and Longitudinal Relaxation Rates between 0.01 and 50 MHz. *Angew. Chem. Int. Ed.* **44**, 2223–2225 (2005).

5. Bolik-Coulon, N., Zachrdla, M., Bouvignies, G., Pelupessy, P. & Ferrage, F. Comprehensive analysis of relaxation decays from high-resolution relaxometry. *J. Magn. Reson.* **355**, 107555 (2023).

6. Pravdivtsev, A. N., Yurkovskaya, A. V., Vieth, H.-M. & Ivanov, K. L. High resolution NMR study of T1 magnetic relaxation dispersion. IV. Proton relaxation in amino acids and Met-enkephalin pentapeptide. *J. Chem. Phys.* **141**, 155101 (2014).

7. Pravdivtsev, A. N., Yurkovskaya, A. V., Petrov, P. A., Vieth, H.-M. & Ivanov, K. L. Analysis of the SABRE (Signal Amplification by Reversible Exchange) Effect at High Magnetic Fields. *Appl. Magn. Reson.* **47**, 711–725 (2016).

8. Roberts, M. F., Cui, Q., Turner, C. J., Case, D. A. & Redfield, A. G. High-Resolution Field-Cycling NMR Studies of a DNA Octamer as a Probe of Phosphodiester Dynamics and Comparison with Computer Simulation. *Biochemistry* **43**, 3637–3650 (2004).

9. Kaptein, R. Chemically induced dynamic nuclear polarization in five alkyl radicals. *Chem. Phys. Lett.* **2**, 261–267 (1968).

10. Ivanov, K. L., Vieth, H.-M., Miesel, K., Yurkovskaya, A. V. & Sagdeev, R. Z. Investigation of the magnetic field dependence of CIDNP in multi-nuclear radical pairs. *Phys. Chem. Chem. Phys.* **5**, 3470–3480 (2003).

11. Pravdivtsev, A. N., Yurkovskaya, A. V., Ivanov, K. L. & Vieth, H.-M. Importance of polarization transfer in reaction products for interpreting and analyzing CIDNP at low magnetic fields. *J. Magn. Reson.* **254**, 35–47 (2015).

12. Grosse, S., Gubaydullin, F., Scheelken, H., Vieth, H.-M. & Yurkovskaya, A. V. Field cycling by fast NMR probe transfer: Design and application in field-dependent CIDNP experiments. *Appl. Magn. Reson.* **17**, 211–225 (1999).

13. Bowers, C. R. & Weitekamp, D. P. Parahydrogen and synthesis allow dramatically enhanced nuclear alignment. *J. Am. Chem. Soc.* **109**, 5541–5542 (1987).

14. Pravica, M. G. & Weitekamp, D. P. Net NMR alignment by adiabatic transport of parahydrogen addition products to high magnetic field. *Chem. Phys. Lett.* **145**, 255–258 (1988).

15. Cavallari, E., Carrera, C., Boi, T., Aime, S. & Reineri, F. Effects of Magnetic Field Cycle on the Polarization Transfer from Parahydrogen to Heteronuclei through Long-Range J-Couplings. *J. Phys. Chem. B* **119**, 10035–10041 (2015).

16. Kaltschnee, L. *et al.* Sensitivity-enhanced magnetic resonance reveals hydrogen intermediates during active [Fe]-hydrogenase catalysis. *Nat. Catal.* **7**, 1417–1429 (2024).

17. Griesinger, C. *et al.* Dynamic nuclear polarization at high magnetic fields in liquids. *Prog. Nucl. Magn. Reson. Spectrosc.* **64**, 4–28 (2012).

18. Kiryutin, A. S. *et al.* A fast field-cycling device for high-resolution NMR: Design and application to spin relaxation and hyperpolarization experiments. *J. Magn. Reson.* **263**, 79–91 (2016).

19. Adams, R. W. *et al.* Reversible Interactions with para-Hydrogen Enhance NMR Sensitivity by Polarization Transfer. *Science* **323**, 1708–1711 (2009).

20. Dücker, E. B., Kuhn, L. T., Münnemann, K. & Griesinger, C. Similarity of SABRE field dependence in chemically different substrates. *J. Magn. Reson.* **214**, 159–165 (2012).

21. Pravdivtsev, A. N., Yurkovskaya, A. V., Vieth, H.-M., Ivanov, K. L. & Kaptein, R. Level Anti-Crossings are a Key Factor for Understanding para-Hydrogen-Induced Hyperpolarization in SABRE Experiments. *ChemPhysChem* **14**, 3327–3331 (2013).

22. Ivanov, K. L., Pravdivtsev, A. N., Yurkovskaya, A. V., Vieth, H.-M. & Kaptein, R. The role of level anti-crossings in nuclear spin hyperpolarization. *Prog. Nucl. Magn. Reson. Spectrosc.* **81**, 1–36 (2014).

23. Colell, J. F. P. *et al.* Generalizing, Extending, and Maximizing Nitrogen-15 Hyperpolarization Induced by Parahydrogen in Reversible Exchange. *J. Phys. Chem. C* **121**, 6626–6634 (2017).

24. Browning, A. *et al.* Spin dynamics of [1,2-$^{13}C_2$]pyruvate hyperpolarization by parahydrogen in reversible exchange at micro Tesla fields. *Phys. Chem. Chem. Phys.* **25**, 16446–16458 (2023).

25. Myers, J. Z. *et al.* Zero to ultralow magnetic field NMR of [1-$^{13}$C]pyruvate and [2-$^{13}$C]pyruvate enabled by SQUID sensors and hyperpolarization. *Phys. Rev. B* **109**, 184443 (2024).

26. TomHon, P., Akeroyd, E., Lehmkuhl, S., Chekmenev, E. Y. & Theis, T. Automated Pneumatic Shuttle for Magnetic Field Cycling and Parahydrogen Hyperpolarized Multidimensional NMR. *J. Magn. Reson.* **312**, 106700 (2020).

27. Pound, R. V. Nuclear Spin Relaxation Times in Single Crystals of LiF. *Phys. Rev.* **81**, 156–156 (1951).

28. Pileio, G., Carravetta, M. & Levitt, M. H. Extremely Low-Frequency Spectroscopy in Low-Field Nuclear Magnetic Resonance. *Phys. Rev. Lett.* **103**, 083002 (2009).

29. Miéville, P., Jannin, S. & Bodenhausen, G. Relaxometry of insensitive nuclei: Optimizing dissolution dynamic nuclear polarization. *J. Magn. Reson.* **210**, 137–140 (2011).

30. Hall, A. M. R., Cartlidge, T. A. A. & Pileio, G. A temperature-controlled sample shuttle for field-cycling NMR. *J. Magn. Reson.* **317**, 106778 (2020).

31. Villanueva-Garibay, J. A. *et al.* A fast sample shuttle to couple high and low magnetic fields. Applications to high-resolution relaxometry. *Magn. Reson. Discuss.* (2025) doi:10.5194/mr-2024-25.

32. Redfield, A. G. High-resolution NMR field-cycling device for full-range relaxation and structural studies of biopolymers on a shared commercial instrument. *J. Biomol. NMR* **52**, 159–177 (2012).

33. Zhukov, I. V. *et al.* Field-cycling NMR experiments in ultra-wide magnetic field range: relaxation and coherent polarization transfer. *Phys. Chem. Chem. Phys.* **20**, 12396–12405 (2018).

34. Ellermann, F., Saul, P., Hövener, J.-B. & Pravdivtsev, A. N. Modern Manufacturing Enables Magnetic Field Cycling Experiments and Parahydrogen-Induced Hyperpolarization with a Benchtop NMR. *Anal. Chem.* **95**, 6244–6252 (2023).

35. Yang, J., Xin, R., Lehmkuhl, S., Korvink, J. G. & Brandner, J. J. Development of a fully automated workstation for conducting routine SABRE hyperpolarization. *Sci. Rep.* **14**, 21022 (2024).

36. Sheberstov, K. *et al.* Robotic arms for hyperpolarization-enhanced NMR. *J. Magn. Reson. Open* **23**, 100194 (2025).

37. Kimmich, R. & Anoardo, E. Field-cycling NMR relaxometry. *Prog. Nucl. Magn. Reson. Spectrosc.* **44**, 257–320 (2004).

38. Chattergoon, N., Martínez-Santiesteban, F., Handler, W. B., Ardenkjær-Larsen, J. H. & Scholl, T. J. Field dependence of T1 for hyperpolarized [1-13C]pyruvate. *Contrast Media Mol. Imaging* **8**, 57–62 (2013).

39. Gizatullin, B., Mattea, C. & Stapf, S. Hyperpolarization by DNP and Molecular Dynamics: Eliminating the Radical Contribution in NMR Relaxation Studies. *J. Phys. Chem. B* **123**, 9963–9970 (2019).

40. Peters, J. P. *et al.* Nitrogen-15 dynamic nuclear polarization of nicotinamide derivatives in biocompatible solutions. *Sci. Adv.* **9**, eadd3643 (2023).

41. Peters, J. *et al.* Chemically induced deceleration of nuclear spin relaxation (CIDER) preserves hyperpolarization. *preprint* (2024) doi:10.21203/rs.3.rs-4668036/v1.

42. Hövener, J.-B. 13C spin hyperpolarization by PASADENA : Instrumentation, preparation of magnetic tracers, and NMR spectroscopy and imaging in vivo. (2008). doi:10.11588/heidok.00008912.

43. Hövener, J.-B. *et al.* PASADENA hyperpolarization of 13C biomolecules: equipment design and installation. *Magn. Reson. Mater. Phy.* **22**, 111–121 (2009).

44. Eills, J. *et al.* Combined homogeneous and heterogeneous hydrogenation to yield catalyst-free solutions of parahydrogen-hyperpolarized [1-13C]succinate. *Chem. Commun.* **59**, 9509–9512 (2023).

45. Tayler, M. C. D. & Levitt, M. H. Paramagnetic relaxation of nuclear singlet states. *Phys. Chem. Chem. Phys.* **13**, 9128–9130 (2011).

46. Pravdivtsev, A. N., Yurkovskaya, A. V., Zimmermann, H., Vieth, H.-M. & Ivanov, K. L. Magnetic field dependent long-lived spin states in amino acids and dipeptides. *Phys. Chem. Chem. Phys.* **16**, 7584–7594 (2014).

47. Truong, M. L. *et al.* 15N Hyperpolarization by Reversible Exchange Using SABRE-SHEATH. *J. Phys. Chem. C* **119**, 8786–8797 (2015).

48. TomHon, P. *et al.* Temperature Cycling Enables Efficient 13C SABRE-SHEATH Hyperpolarization and Imaging of [1-13C]-Pyruvate. *J. Am. Chem. Soc.* **144**, 282–287 (2022).

49. Kowalewski, J. & Mäler, L. *Nuclear Spin Relaxation in Liquids: Theory, Experiments, and Applications*. (Taylor&Francis, 2006).

50. Hundshammer, C., Grashei, M., Greiner, A., Glaser, S. J. & Schilling, F. pH Dependence of T1 for 13C-Labelled Small Molecules Commonly Used for Hyperpolarized Magnetic Resonance Imaging. *ChemPhysChem* **20**, 798–802 (2019).

51. Knecht, S., Pravdivtsev, A. N., Hövener, J.-B., Yurkovskaya, A. V. & Ivanov, K. L. Quantitative description of the SABRE process: rigorous consideration of spin dynamics and chemical exchange. *RSC Adv.* **6**, 24470–24477 (2016).

52. Eriksson, S. L., Lindale, J. R., Li, X. & Warren, W. S. Improving SABRE hyperpolarization with highly nonintuitive pulse sequences: Moving beyond avoided crossings to describe dynamics. *Sci. Adv.* **8**, eabl3708 (2022).

53. MacCulloch, K. *et al.* Hyperpolarization of common antifungal agents with SABRE. *Magn. Reson. Chem.* **59**, 1225–1235 (2021).

54. Barskiy, D. A. *et al.* The Feasibility of Formation and Kinetics of NMR Signal Amplification by Reversible Exchange (SABRE) at High Magnetic Field (9.4 T). *J. Am. Chem. Soc.* **136**, 3322–3325 (2014).

55. Pravdivtsev, A. N., Yurkovskaya, A. V., Zimmermann, H., Vieth, H.-M. & Ivanov, K. L. Transfer of SABRE-derived hyperpolarization to spin-1/2 heteronuclei. *RSC Adv.* **5**, 63615–63623 (2015).

56. Assaf, C. D. *et al.* Analysis of Chemical Exchange in Iridium N-Heterocyclic Carbene Complexes Using Heteronuclear Parahydrogen-Enhanced NMR. *Commun. Chem.* **7**, 286 (2024).

57. Pravdivtsev, A. N. *et al.* Coherent Evolution of Signal Amplification by Reversible Exchange in Two Alternating Fields (alt-SABRE). *ChemPhysChem* **22**, 2381 (2021).

58. J. Tickner, B. & B. Duckett, S. Iridium trihydride and tetrahydride complexes and their role in catalytic polarisation transfer from para hydrogen to pyruvate. *Chem. Sci.* **16**, 1396–1404 (2025).

59. Assaf, C. D. *et al.* J Coupling Constants of <1 Hz Enable 13C Hyperpolarization of Pyruvate via Reversible Exchange of Parahydrogen. *J. Phys. Chem. Lett.* 1195–1203 (2024) doi:10.1021/acs.jpclett.3c02980.

60. Morris, G. A. & Freeman, R. Enhancement of nuclear magnetic resonance signals by polarization transfer. *J. Am. Chem. Soc.* **101**, 760–762 (1979).

61. Cowley, M. J. *et al.* Iridium N-Heterocyclic Carbene Complexes as Efficient Catalysts for Magnetization Transfer from para-Hydrogen. *J. Am. Chem. Soc.* **133**, 6134–6137 (2011).

62. Schmidt, A. B. *et al.* Selective excitation of hydrogen doubles the yield and improves the robustness of parahydrogen-induced polarization of low-γ nuclei. *Phys. Chem. Chem. Phys.* **23**, 26645–26652 (2021).

63. Pravdivtsev, A. N., Yurkovskaya, A. V., Zimmermann, H., Vieth, H.-M. & Ivanov, K. L. Enhancing NMR of insensitive nuclei by transfer of SABRE spin hyperpolarization. *Chem. Phys. Lett.* **661**, 77–82 (2016).

64. Salnikov, O. G. *et al.* Modeling Ligand Exchange Kinetics in Iridium Complexes Catalyzing SABRE Nuclear Spin Hyperpolarization. *Anal. Chem.* **96**, 11790–11799 (2024).

65. Chiavazza, E. *et al.* Earth's magnetic field enabled scalar coupling relaxation of 13C nuclei bound to fast-relaxing quadrupolar 14N in amide groups. *J. Magn. Reson.* **227**, 35–38 (2013).

66. Kharkov, B. *et al.* Weak nuclear spin singlet relaxation mechanisms revealed by experiment and computation. *Phys. Chem. Chem. Phys.* **24**, 7531–7538 (2022).